\documentclass{article}
\usepackage{latexsym}

\newcommand{\RR}{\mbox{${\rm \:  R\!\!\!\! I
\;\;}$}}

\newcommand{\qed}{\hfill $\Box$ \vskip 2ex}
\newcommand{\vs}{\vspace{0.25cm}}

\newtheorem{theorem}{Theorem}
\newtheorem{itlemma}{Lemma}[section]
\newtheorem{itproposition}[itlemma]{Proposition}
\newtheorem{itcorollary}[itlemma]{Corollary}
\newtheorem{itremark}[itlemma]{Remark}
\newtheorem{itremarks}[itlemma]{Remarks}
\newtheorem{itdefinition}[itlemma]{Definition}
\newtheorem{itexample}[itlemma]{Example}

\newenvironment{lemma}{\begin{itlemma}\rm}{\end{itlemma}} 
\newenvironment{remark}{\begin{itremark}\rm}{\end{itremark}} 
\newenvironment{remarks}{\begin{itremarks} \rm}{\end{itremarks}}
\newenvironment{corollary}{\begin{itcorollary}\rm}{\end{itcorollary}}
\newenvironment{proposition}{\begin{itproposition}\rm}{\end{itproposition}}
\newenvironment{definition}{\begin{itdefinition}\rm}{\end{itdefinition}}
\newenvironment{example}{\begin{itexample}\rm}{\end{itexample}}
\newenvironment{fact}{\noindent {\em Fact}. \ \ }{\hfill \medskip}
\newenvironment{proof}{\noindent {\em Proof}.\ \
}{\hspace*{\fill}$\Box$\medskip}
\newenvironment{claim}{\noindent {\em Claim}. \ \ }{\hfill \medskip}
\newcommand{\be}[1]{\begin{equation}\label{#1}}
\newcommand{\ee}{\end{equation}}
\newcommand{\bl}[1]{\begin{lemma}\label{#1}}
\newcommand{\br}[1]{\begin{remark}\label{#1}}
\newcommand{\brs}[1]{\begin{remarks}\label{#1}}

\newcommand{\bt}[1]{\begin{theorem}\label{#1}}
\newcommand{\bd}[1]{\begin{definition}\label{#1}}
\newcommand{\bp}[1]{\begin{proposition}\label{#1}}
\newcommand{\bc}[1]{\begin{corollary}\label{#1}}
\newcommand{\bfact}[1]{\begin{fact}\label{#1}}
\newcommand{\bex}[1]{\begin{example}\label{#1}}
\newcommand{\ec}{\end{corollary}}
\newcommand{\efact}{\end{fact}}
\newcommand{\eex}{\end{example}}
\newcommand{\el}{\end{lemma}}
\newcommand{\er}{\end{remark}}
\newcommand{\ers}{\end{remarks}}
\newcommand{\et}{\end{theorem}}
\newcommand{\ed}{\end{definition}}
\newcommand{\ep}{\end{proposition}}
\newcommand{\epr}{\end{proof}}
\newcommand{\bpr}{\begin{proof}}
\newcommand{\bcl}{\begin{claim}}
\newcommand{\ecl}{\end{claim}}

\newcommand{\bi}{\begin{itemize}}
\newcommand{\ei}{\end{itemize}}
\newcommand{\ben}{\begin{enumerate}}
\newcommand{\een}{\end{enumerate}}
\newcommand{\text}[1]{\hbox{\rm \ #1\ \/}}

\begin{document}

\begin{center}

{\Large Analysis and identification of quantum dynamics using Lie
algebra homomorphisms   and Cartan decompositions}

\vs

\vs

Francesca Albertini\footnote{Department of Pure and Applied
Mathematics, University of Padova, Via Belzoni 7, 35100 Padova,
Italy, e-mail:albertin@math.unipd.it, Tel: +39-049-8275966} and
Domenico D'Alessandro\footnote{Department of Mathematics, Iowa State
University, Ames, IA 50011, U.S.A. e-mail: daless@iastate.edu, Tel.
+1-515-294-8130}

\end{center}

\begin{abstract}

In this paper, we consider the problem of model equivalence for
quantum systems. Two models are said to be (input-output) equivalent
if they give the same output for every admissible input. In the case
of quantum systems, the output is the expectation value of a given
observable or, more in general, a probability distribution for the
result of a quantum measurement.  We link the input-output
equivalence of two  models to the existence of a homomorphism of the
underlying Lie algebra. In several cases, a Cartan decomposition of
the Lie algebra $su(n)$ is useful to find such a homomorphism and to
determine the classes of equivalent models. We consider in detail
the important  cases of two level systems with a Cartan structure
and  of spin networks. In the latter case, complete results are
given generalizing previous results to the case of networks of spin
particles with any value of the spin. In treating this problem, we
prove some instrumental results on the subalgebras of $su(n)$ which
are of independent interest.

\end{abstract}

\vs


\vs

\noindent{\bf Running Title}: Analysis and identification of quantum
dynamics.

\section{Introduction}
\label{INTRO}

Dynamical models of quantum systems have been recently the subject
of investigation, concerning their  structural properties, by  use
of methods of control theory. Appropriate definitions of
controllability and observability of quantum systems have been given
and practical conditions to check these properties have been
proposed (see e.g. \cite{confraIEEE}, \cite{ioobs},
\cite{HuangTarn}, \cite{RamaK}). In many cases, the tools used are
the ones of Lie algebra and Lie group theory. Information on the
properties  of the dynamics is obtained by a study of the structure
of a Lie algebra associated to the system and how this relates to
the particular equations at hand. This geometric approach has proved
useful not only to analyze the dynamics but also to design control
laws. This approach can also be used  to study problems of parameter
identification of quantum systems and this is the subject of the
present paper. In particular, the problem we shall study is the
classification of models of quantum systems whose behavior cannot be
distinguished by an external observer. We shall call these models
{\it (input-output) equivalent}. This problem is motivated by
several experimental scenarios. In particular consider a molecule
which is a network of particles with spin with  all the other
degrees of freedom neglected. A model Hamiltonian is associated to
this system in which parameters modeling the interaction between
particles as well as the interaction with an external
electro-magnetic field are unknown. Also,  the initial state of the
system might be unknown. In experimental scenarios such as Nuclear
Magnetic Resonance and Electron Paramagnetic Resonance, it is
possible to drive the system with a magnetic field and measure the
expectation value of a given observable,  for example the total spin
in a given direction. The question of fundamental and practical
importance is to what extent, with this type of experiments, it is
possible to distinguish between different models. As we shall see in
this paper, this  question is related to the existence of a
particular Lie algebra homomorphism which relates  the equations of
the two models.

\vs

The main results of this paper are the solution of the model
equivalence problem for a class of two level systems in Theorem
\ref{TLC}, a number of auxiliary results (Theorems
\ref{Teorema1}-\ref{Teorem3}) on the structure of the Lie algebra
$su(n)$ and its subalgebras and Theorem \ref{main} where we
completely solve the problem of characterizing equivalent models for
networks of spin. The latter  result generalizes results previously
obtained in \cite{LAA2005} and \cite{ioIEEE} , which were proven
only for networks of spin $\frac{1}{2}$ and $1$'s, to networks of
interacting spins  of  any value and where the spin itself is an
unknown  parameter to be identified. A further motivation for this
research can be found in \cite{Luban} where it was shown that
thermodynamic methods commonly used to identify the parameters of
spin networks such as in molecular magnets \cite{MM1}, \cite{MM2}
are not always adequate. The generalization  presented in this paper is obtained
through a Cartan decomposition technique recently presented in
\cite{OddEven} which helps determining the homomorphism between
equivalent models in the form of   a Cartan involution.

\vs

The paper is organized as follows. In Section \ref{PME}, we describe
the problem of model equivalence for quantum systems. In Section
\ref{GSC}, we link the equivalence of two models to the existence of
an appropriate Lie algebra homomorphism.  In several cases the
structure of the dynamics is related to a Cartan decomposition of
$su(n)$ and suggests the form of such a homomorphism as well as of the
classes of equivalent models. We give a two level example  in
Section \ref{MET} and treat the case of   general spin networks in
Section \ref{MES}. Instrumental to the solution of the model equivalence
problem for spin networks are some results of independent interest
concerning the existence of  subalgebras of $su(n)$ with specific
features. The proofs of these results are presented in Section
\ref{Proofs}. Concluding remarks  are given in Section \ref{CONCLU}.

\section{The problem of model equivalence for quantum systems}
\label{PME}

Consider a model Hamiltonian for a quantum system, $H(t):=H(u(t))$,
where, in a semiclassical description,  the dependence on time is
due to the interaction with external fields,  $u:=u(t)$,  which play
the role of {\it controls}. The evolution of the state of the
system, described by a density matrix $\rho:=\rho(t)$, is
determined, other than by $H$,   by the initial state
$\rho(0)=\rho_0$. In particular, $\rho$ is the solution of the
Liouville's equation \be{Liouville1} \dot \rho=[-iH, \rho],  \ee
with initial condition $\rho(0)=\rho_0$. For this system, we measure
an observable $S$. Considering, for simplicity, a Von
Neumann-L\"uders measurement\footnote{Natural extensions can be made
to  general measurements \cite{Petruccione} for the related issue of
observability of quantum systems.}, writing $S$ in terms of
 orthogonal projections \be{Spro} S:=\sum_j \lambda_j \Pi_j,   \ee
 the probability of having a result $\lambda_j$, when the state is
 $\rho$, is given by
 \be{probabilita}
 P_j:=Tr(\Pi_j \rho).
 \ee
 As the probabilities
 $P_j$ are the only information that can be gathered by an external
 observer, it is motivated to ask what classes of models $\{ H(u),
 \rho_0 \}$ will give the same probabilities, for any functional
 form  of the control $u$. In other terms, we ask what classes of
 models are indistinguishable by experiments that involve driving
 the system with  controls, in a given set of functions,
  and measuring a given observable. These models will be called
  {\it(Input-Output) Equivalent}.

  It is appropriate to treat the
  case where the {\it expectation value} of the measurable $S$, i.e.
  the `output',
\be{expval} y:=Tr(S\rho),
 \ee
is measured. Not only this is the case in several experimental
situations, such as nuclear magnetic resonance, but it is not a
significant restriction as compared to the case where the
probabilities (\ref{probabilita})  are considered. As the structure
of the output (\ref{expval}) is the same as the one of the outputs
(\ref{probabilita}), the passage {}from the treatment for the
expectation value to the one for probabilities corresponds to
extending a single output treatment to a multiple output treatment.
This  can be accomplished without difficulties.

\vs

In order to render the problem of characterizing  the classes of
equivalent models  treatable, we need to assume some structure on
the Hamiltonian $H$. This corresponds to the passage {}from {\it
unstructured uncertainty} to {\it parametric uncertainty} often
discussed in identification theory (see e.g. \cite{SastryBodson}).
In particular, it is often the case that the Hamiltonian $H=H(u)$
has the bilinear form \be{bilinear} H\, :=\,
H_0+\sum_{j=1}^mH_ju_j(t), \ee for some control functions
$u_1,...,u_m$, and {\it internal Hamiltonian} $H_0$ and {\it
interaction Hamiltonians} $H_j$'s, $j=1,...,m$. These can be
considered as Hermitian matrices of dimension $n$, i.e. elements of
$isu(n)$, where $n$ is the dimension of the system, assumed finite.
Moreover, in many cases, $H_0$ and the $H_j$'s belong to two
orthogonal complementary subspaces of $isu(n)$ corresponding to  a
Cartan decomposition of $su(n)$ \cite{Helgason}. These are two
subspaces $i \cal K$ and $i \cal P$, such that the subspaces of
$su(n)$, $\cal K$ and $\cal P$, satisfy the commutation relations
\be{commutat} [{\cal K}, {\cal K}] \subseteq {\cal K}, \qquad [{\cal
K}, {\cal P}] \subseteq {\cal P}, \qquad [{\cal P}, {\cal P}]
\subseteq {\cal K}. \ee If the system is a multipartite system,
every $H_j$ is a linear combination of Hamiltonians modeling the
interaction of each individual system with the external field. In
matrix notation, $H_j$ is a linear combination of elements of the
type   ${\bf 1} \otimes {\bf 1} \otimes \cdot \cdot \cdot \otimes
{\bf 1} \otimes L \otimes {\bf 1} \otimes \cdot \cdot \cdot \otimes
{\bf 1}$, where $L$ is an Hermitian matrix of appropriate dimensions
and all the other places are occupied by identities $\bf 1$. Also,
$H_0$ is very often a linear combination of elements modeling the
interaction between two subsystems, which can be written as tensor
products of matrices equal to the identity except in two locations.
In these cases the relevant  Cartan decomposition for  the
Hamiltonian (\ref{bilinear}) can often be  chosen of the {\it
odd-even type}  described in the recent paper \cite{OddEven}. Also
if $S$ is a sum of observables on each individual subsystem, i.e.
total angular momentum (see e.g. \cite{Sakurai}), it can always be
written as sum of tensor products all equal to the identity except
in one position. In these cases $iS$,   belongs to one of the
subspaces of the Cartan decomposition\footnote{Notice that the
situation may be different if we consider the case of a single
output given by the expectation value (\ref{expval}) and the case of
several outputs given by the probabilities in (\ref{probabilita}).}.

\vs

In the following, we shall consider, as standing assumption, only
finite dimensionality of the Hamiltonian $H$ and the bilinear form
 (\ref{bilinear}) and will make precise the assumptions on the
Cartan structure of the Hamiltonian when needed.

\section{Model equivalence and Lie algebra homomorphisms}
\label{GSC}

Consider two models with Hamiltonian of the form (\ref{bilinear})
and output of the form (\ref{expval}) \be{Hamilsys1} \dot
\rho=[-i(H_{0}+\sum_{j=1}^{m} H_{j} u_{j}), \rho], \quad
\rho(0)=\rho_{0}, \quad y=Tr(S\rho), \ee
\be{Hamilsys2} \dot
\rho'=[-i(H_{0}'+\sum_{j=1}^{m} H_{j}' u_{j}), \rho'], \quad
\rho'(0)=\rho_{0}', \quad y'=Tr(S'\rho'). \ee The following theorem
links the existence of an appropriate Lie algebra homomorphism to
the equivalence of the two models.

\bt{THMGEN} Let $n$ and $n'$ be the dimensions of the two models
(\ref{Hamilsys1}), (\ref{Hamilsys2}), respectively. Let $\phi$ be a
homomorphism, $\phi : u(n) \rightarrow u(n')$, and $\phi^{*}$ its dual
with respect to the standard inner product $<A,B>:=tr(AB^{*})$.
Assume
\be{condi11}
-iH_{0}'=\phi(-iH_{0}), \qquad -iH_{j}'=\phi(-iH_{j}), \qquad
\phi^{*}(iS')=iS.
\ee
Then if $i\rho_{0}'=\phi(i\rho_{0})$ the models
are equivalent. Viceversa, if the models are equivalent and
(\ref{Hamilsys2}) is observable, then
\be{statieq}
i\rho_{0}'=\phi(i\rho_{0}).
\ee
\et

\bpr Multiply (\ref{Hamilsys1}) and (\ref{Hamilsys2}) by $i$ and
then apply $\phi$ to the equation obtained {}from (\ref{Hamilsys1}).
Combining the two resulting equations, we obtain \be{combinatio}
\frac{d}{dt}(i \rho'-\phi(i\rho))= [ \phi(-i
H_{0})+\sum_{j}\phi(-iH_{j})u_{j}, i \rho'-\phi(i\rho)]. \ee Now, if
(\ref{statieq}) is verified, then $i\rho'(t)= \phi(i\rho(t))$, for
every $t$ and for every control. Therefore we have \be{Eqqq}
Tr(S'\rho')=Tr(-iS'
i(\rho'))=Tr(-iS'\phi(i\rho))=Tr(\phi^{*}(-iS')i\rho)=Tr(S\rho), \ee
and the two models are equivalent. Viceversa, assume the two models
are equivalent. {}From (\ref{Eqqq}), we have \be{coseq} Tr(iS'
(i\rho'-\phi(i \rho))(t))=0, \ee for every $t$. Writing the solution
of (\ref{combinatio}) as $(i\rho'-\phi(i\rho)(t)=X
(i\rho'-\phi(i\rho))(0) X^{*}$, where $X$ is the solution of the
(Schr\"odinger) operator equation $\dot X=( \phi(-i
H_{0})+\sum_{j}\phi(-iH_{j})u_{j})X,$   $X(0)={\bf 1}$, we have
\be{coseq1} Tr(X^*iS'X (i\rho'_{0}-\phi(i \rho_{0})))=0. \ee As the
system (\ref{Hamilsys2}) is observable, we have that  $X^*iS'X$ span
all of $su(n')$, which implies $i\rho'_{0}=\phi(i \rho_{0})$. \epr

As we shall show in the remainder of the paper (cf. also
\cite{LAA2005}),  it is possible for cases of physical interest to
give a stronger version of Theorem \ref{THMGEN}. In particular, it
is possible to show that the existence of a homomorphism $\phi$
satisfying (\ref{condi11}) is also necessary for equivalence of two
models. This way, we can characterize all the classes of equivalent
models in terms of homomorphisms. We shall do this for a two level
example in the next section and general spin networks in Section \ref{MES}.
In both cases we exploit a Cartan decomposition underlying the
dynamics of the models. In general, more structure will have  to be
assumed to avoid trivial cases. For example, if $S=S'$ is a scalar
matrix, then every two models are equivalent.  To avoid this case, a
reasonable
 extra assumption is the observability of the two models.
 Also, we need to  assume
 that the initial states are not  both perfect mixtures,
 otherwise, with $S=S'$,  the output for any
 two equivalent models will be the same,
 independently of the dynamics. Moreover $-iH_{j}$ and $-iH_{j}'$,
 $j=0,\ldots,m$,
 may be in general assumed  traceless, as the trace only adds
  an extra common phase factor to the dynamics,  which cannot
 be detected.

\vs

\section{Model equivalence of two level systems}
\label{MET}

Consider  a spin $\frac{1}{2}$ particle which is driven by an
electro-magnetic control field along the $z$ axis, interacts with a
constant unknown magnetic field  along a (unknown) direction in the
$x-y$ plane and has unknown initial state. The practical question is
to what extent, by driving the system with the control field and
measuring the average value of the spin magnetization in the $z$
direction, it is possible to obtain information about the unknown
parameters of the system. This type of model has a Cartan structure
which is shared by several other models of physical interest and is
instrumental in finding a homomorphism between equivalent models. We
describe this below.

\vs

The Lie algebra $su(2)$, which is the relevant Lie algebra in the
two level case, has, up to conjugacy,  only one Cartan decomposition
(or in other terms all the types of decompositions coincide in this
case) which corresponds to the classical Euler decomposition of the
Lie group $SU(2)$ \cite{Helgason}. This extends to a  decomposition
of $u(2)$ which can always be written, as \be{decou2} u(2)={\cal K}
\oplus {\cal P}.   \ee Here $\cal K$ and $\cal P$ satisfy the
commutation relations in (\ref{commutat}) and are given, up to
conjugacy,  by \be{calKa} {\cal K}:=span \{ i \sigma_z \}, \qquad
{\cal P}:= span \{ i\sigma_x, i\sigma_y, i{\bf 1}_{2 \times 2} \},
\ee where ${\bf 1}_{2 \times 2}$ is the $2\times 2$ identity matrix
and $\sigma_x$, $\sigma_y$ and $\sigma_z$ are the Pauli matrices
\be{Paulim} \sigma_x:=\frac{1}{2} \pmatrix{0 & 1 \cr 1 & 0}, \quad
\sigma_y:= \frac{1}{2} \pmatrix{0 & i \cr -i & 0}, \quad \sigma_z:=
\frac{1}{2} \pmatrix{1 & 0 \cr 0 & -1}.\ee The dynamical and output
equation, for the above model of a spin $\frac{1}{2}$ particle in an
electro-magnetic field,  can be written as \be{1equi1} \dot \rho=[A+
i\sigma_z u(t), \rho], \quad y=Tr(\sigma_z \rho), \quad
\rho(0)=\rho_0,   \ee where $\rho_0$ is an unknown initial density
matrix and $A:=xi\sigma_x+yi \sigma_y$, with $x$ and $y$ unknown.
This model has a {\it Cartan structure} in that $A$ is in  ${\cal
P}$ and $i \sigma_z$ (the control and observation part) is in ${\cal
K}$, with $\cal K$ and $\cal P$ defined in (\ref{calKa}). We assume
$x^2+y^2 \not= 0$ which implies controllability and therefore
observability \cite{ioobs} for this model. The following result
characterizes all the classes of equivalent models in terms of Lie
algebra homomorphisms.

 \bt{TLC} Consider two  models
\be{equi1} \dot \rho=[A+ i\sigma_z u(t), \rho], \quad y=Tr(\sigma_z
\rho), \quad \rho(0)=\rho_0,   \ee \be{equi2} \dot {\rho}'=[A'+
i\sigma_z u(t), \rho], \quad y=Tr(\sigma_z \rho'), \quad
\rho'(0)=\rho_0',\ee with $\rho_0$ and $\rho_0'$ not both equal to
scalar
 matrices (representing perfect mixtures) and $A$ and $A'$
 given by
\be{Acomp} A:=x i \sigma_x + y i \sigma_y,    \ \ \text{ and } \  \
A':=x'i \sigma_x+y'i\sigma_y. \ee Assume \be{contrrr} x^2+y^2 \not=0
\\ \text{ and}  \\ x'^2 +y'^2 \not=0.
\ee
Then the two models are equivalent if and only if there exists
an automorphism $\phi: u(2) \rightarrow u(2)$ with
 \be{condiz1}
 \phi^*(i
 \sigma_z)=i \sigma_z
 \ee and
 \be{repeatconj} A'=\phi(A), \quad \phi(i\sigma_{z})=i\sigma_{z},
 \quad  i\rho_0'=\phi(i \rho_0).
 \ee
\et

\bpr  It is clear that if the automorphism $\phi$ exists, satisfying
(\ref{condiz1}) (\ref{repeatconj}), the two models are equivalent.
This follows {}from a direct application of Theorem \ref{THMGEN}. To
prove the opposite, first notice that, {}from the equivalence
assumption, we have \be{tryeq} y(t):=Tr(\sigma_z
\rho(t))=Tr(\sigma_z \rho'(t)):=y'(t), \ee for every $t \geq 0$ and
every admissible control.

\vs

We consider an inner automorphism $\phi$ of
the following type \be{fidef}
\phi(L):=e^{-i\alpha \sigma_z} L e^{i\alpha \sigma_z}, \quad L \in
u(n), \ee
as $\alpha$ varies in $\RR$.

Clearly (\ref{condiz1}) and the second one of (\ref{repeatconj})
are  verified for any $\alpha\in \RR$.
Moreover \be{Atransfor} \phi(A)=\bar x i \sigma_x+\bar y i \sigma_y,
\ee with \be{plmnb} \pmatrix{\bar x \cr \bar y}= K_{\alpha}
\pmatrix{x \cr y}, \ee and \be{kappa}
K_{\alpha}:=\pmatrix{cos(\alpha) &  sin(\alpha) \cr - sin(\alpha) &
cos(\alpha)}. \ee Also, if we write \be{iro}
\begin{array}{l}
    i \rho_0 :=\rho_x i \sigma_x + \rho_y  i \sigma_y+
\rho_z  i \sigma_z +\frac{1}{2} i {\bf 1},
\\   \\
i \rho'_0 :=\rho'_x i \sigma_x + \rho'_y  i \sigma_y+
\rho'_z  i \sigma_z +\frac{1}{2} i {\bf 1} \end{array}
\ee we have \be{hhhhj}
\phi(i \rho_0 )= \bar \rho_x i \sigma_x + \bar \rho_y i
\sigma_y+ \bar \rho_z i \sigma_z +\frac{1}{2} i {\bf 1}, \ee with
\be{plmnbbis} \pmatrix{\bar \rho_x \cr \bar \rho_y}= K_{\alpha}
\pmatrix{\rho_{x} \cr \rho_y}.  \ee
Using the equivalence assumption (\ref{tryeq}) at $t=0$ we
immediately obtain \be{RZbis} \rho_z=\rho_z'. \ee Moreover,
differentiating (\ref{tryeq}) using the dynamical equations
(\ref{equi1}) (\ref{equi2}), we obtain \be{ggghg} Tr(\rho[\sigma_z,
A])=  Tr(\rho'[\sigma_z, A']).\ee Writing this at time $t=0$ and
using the definitions (\ref{Acomp}) and (\ref{iro}) along with the
commutation relation for the Pauli matrices \be{Commurel} [i
\sigma_x, i \sigma_y]=i \sigma_z, \qquad  [i \sigma_y, i \sigma_z]=i
\sigma_x, \qquad  [i \sigma_z, i \sigma_x]=i \sigma_y,   \ee we obtain

\be{obt1} \rho_y x-\rho_x y=\rho_y'x'-\rho_x'y'.  \ee
Differentiating (\ref{ggghg}) and using the fact that the resulting
equation has to be valid for every value of the control, we obtain
the two equations \be{e1} Tr(\sigma_z[A,[A,\rho]])=
Tr(\sigma_z[A',[A',\rho']]), \ee and \be{e2} Tr(i
\sigma_z[A,[\sigma_z,\rho]])=Tr(i \sigma_z[A',[\sigma_z,\rho']]).
\ee

{}From equation (\ref{e2}), as for equation (\ref{obt1}), we obtain

\be{obt2} x \rho_x + y \rho_y= x' \rho_x'+ y' \rho_y'. \ee

{}From equation (\ref{e1}), we obtain

\be{obt3} (x^2+y^2) Tr(\sigma_z \rho)=(x'^2+y'^2) Tr(\sigma_z
\rho').   \ee Using the fact that $Tr(\sigma_z \rho)$ is not
always zero (because of the controllability condition
(\ref{contrrr})) \footnote{{}From controllability
(\ref{contrrr}), we cannot have \be{cannot} Tr(\sigma_z
\rho(t)) \equiv Tr(\sigma_z \rho'(t))\equiv 0, \ee for every control. This
would mean that, for every reachable evolution operator $X$,
solution of the (Schr\"odinger) operator equation $\dot X=(A+i
\sigma_{z}u) X$,   $X^*
\sigma_z X$ would be orthogonal to $\rho_0$. However, because
 of controllability $X$ may attain all the values in $SU(2)$ and
 therefore $X^* \sigma_z X$ span, as $X$ varies,  all of
$isu(2)$. Therefore,  $X^* \sigma_z X$ is always orthogonal to
$\rho_0$ only if $\rho_0$ is a multiple of the identity, which we
have excluded.} and equation (\ref{tryeq}), we have  \be{klm}
x^2+y^2=x'^2+y'^2.  \ee Therefore,  for some $\alpha$, we can write
\be{tty} \pmatrix{x' \cr y'}=K_{\alpha} \pmatrix{x \cr y}, \ee with
$K_{\alpha}$ in (\ref{kappa}), and this, compared with (\ref{plmnb})
and (\ref{Atransfor}),  gives the first one of (\ref{repeatconj}).
To obtain the third  one (with the same $\phi$), we recall
(\ref{contrrr})  that $x^2+y^2 \not=0$. Letting $J:=\pmatrix{0 & 1
\cr -1 & 0}$ and using  (\ref{tty}), we can write (\ref{obt1}) and
(\ref{obt2}), respectively,  as

\be{obt1bis} [x,y] J [\rho_{x},\rho_{y}]^{T}=[x,y] K_{\alpha}^T J
[\rho_{x}',\rho_{y}']^{T},  \ee

\be{obt2bis}  [x,y]
[\rho_{x},\rho_{y}]^{T}=[x,y]K_{\alpha}^T[\rho_{x}',\rho_{y}']^{T}.
\ee Since $K_{\alpha}^T$ commutes with $J$, we can write these as

\be{compacta} \pmatrix{[x,y]J \cr [x,y]}[\rho_{x},\rho_{y}]^{T}=
 \pmatrix{[x,y]J \cr [x,y]} K_{\alpha}^T [\rho_{x}',\rho_{y}']^{T}. \ee Since
$x^2+y^2=-det  \pmatrix{[x,y]J \cr [x,y]}\not= 0$, we can
write
\be{tytytyt} [\rho_{x}',\rho_{y}']^{T}=K_{\alpha}[\rho_{x},\rho_{y}]^{T},  \ee
and therefore
\be{olpo} [\rho_x', \rho_y']= [\bar \rho_x, \bar \rho_y], \ee
which along with $\rho_z'=\rho_z$ gives \be{gj} i\rho'=\phi(i\rho).
\ee This concludes the proof of the Theorem. \epr


\section{Model equivalence of spin networks}
\label{MES}

\subsection{Set of models of spin networks}

We consider a network of $n$ particles with spin  that interact
according to  Heisenberg interaction. In particular, we denote  the
spin of the $j^{th}$ particle by $l_j$ and by $N_j:=2l_j+1$ the
dimension of the Hilbert space for the state of the $j^{th}$
particle. The dimension of the Hilbert state space associated to the
entire network is $N:=\prod_{j=1}^n N_j$. The class of Hamiltonians
we consider are of the form

\be{dinamica}
H(t):= i(A+B_xu_x(t)+B_yu_y(t)+B_zu_z(t)),   \ee
 where $A$,  modeling  the Heisenberg interaction among the
 particles, and $B_{x,y,z}$, modeling the interaction with external fields,
  are given by
\be{dinamica1}
\begin{array}{ccl}
A &:= & -i \sum_{k<l, k,l=1}^n J_{kl}(I_{kx,lx}+ I_{ky,ly}+
I_{kz,lz}).  \\
   & & \\
B_{v}&:= & -i (\sum_{k=1}^n \gamma_k I_{kv}), \ \ \text{ for }
v=x,\, y, \text{ or } z,  \end{array} \ee respectively. Here and in
the following we denote by $I_{k_1v_1,...,k_rv_r}$, for $1\leq k_1
<\cdots<k_r\leq n$ and $v_j\in \{x,\, y,\, z \}$, the $N\times N$
matrix which is the Kronecker product of $n$ matrices where in the
$j^{th}$ position we have the $N_j\times N_j$ identity if $j \not\in
\{ k_1,\ldots,k_r\}$, while if $j=k_s$ we have the $N_j\times N_j$
representation of the $v_{s}$ component of spin angular momentum for
a particle with spin $l_{j}$. Such matrices are given by the Pauli
matrices (\ref{Paulim}) in the case where $l_{j}=\frac{1}{2}$ and
can be calculated for every value of the spin (see e.g.
\cite{Sakurai} Section 3.5). With some abuse of notation, we shall
continue denoting these matrices by $\sigma_{x}$, $\sigma_{y}$ and
$\sigma_{z}$, without explicit reference to the value of the spin.
These matrices have several properties we shall use in the
following. In particular, they satisfy the commutation relations
(\ref{Commurel}) and \be{nu4}
\sigma_{x}^2+\sigma_{y}^{2}+\sigma_{z}^{2}=l_{j}(l_{j}+1) {\bf
1}_{N_{j}\times N_{j}} \ee (see e.g. formula (3.5.34a) in
\cite{Sakurai}).

The real scalar parameter $J_{kl}$ in (\ref{dinamica1}) is   the
exchange constant between particle $k$ and particle $l$ and the real
scalar parameter $\gamma_k$ is the gyromagnetic ratio of particle
$k$. We assume that the spins  of the  network have all  non-zero
and different gyromagnetic ratios.  We can associate a graph to the
model, where each node represents a particle and an edge connects
two nodes if and only if the corresponding exchange constant is
different {}from zero. It is not difficult to see  that
 if the model is
controllable then, necessarily, this graph is connected. Moreover,
controllability implies observability for every  output of the form
(\ref{expval}) where $S$ is a non scalar matrix \cite{ioobs}. In our
case, we assume to measure the expectation values of the total
magnetization in the $x$, $y$, and $z$ direction, given as in
(\ref{expval}) where $S$ is one of   the matrices: \be{expvalmag}
S_{v}:= \sum_{k=1}^{n} I_{k{v}},  \ \text{ with } v\in \{x,\, y, \,z
\}. \ee A model of the type above described will be denoted by
$\Sigma:= \Sigma(n,l_{j},J_{kl},\gamma_{k},\rho_{0})$, where   the
parameters $n,l_{j},J_{kl},\gamma_{k},\rho_{0}$, which determine the
model, are unknown. We will assume to have two controllable models
 $\Sigma$ and $\Sigma':=\Sigma'(n',l_{j}',J_{kl}',
\gamma_{k}',\rho_{0}')$ which satisfy the previous requirements and
we look for necessary and sufficient conditions for these two models
to be equivalent. We shall mark with a prime, $'$, all the
quantities concerning the system $\Sigma'$.

\subsection{Relevant homomorphism of $su(n)$}


In \cite{OddEven} a method was described to construct a Cartan
decomposition of the Lie algebra $su(N)$ for a multipartite system,
starting {}from   decompositions of the Lie algebras
$su(N_j)$ associated to  the
single subsystems, each of dimension $N_j$, with $N:=\prod_{j=1}^n
N_j$. In particular, we have the following result. \bt{oeth}
(\cite{OddEven}, Section 5) Consider a multi partite system with $n$
subsystems of dimensions $N_{1},\ldots,N_{n}$. Consider the Lie
algebra $u(N_{j})$ related to the $j$-th subsystem and a Cartan
decomposition \be{63} u(N_{j})= {\cal{K}}_{j} \oplus {\cal{P}}_{j},
\ee of the type {\bf{AI}} or {\bf{AII}} \footnote{In a decomposition
{\bf{AI}} ${\cal{K}}_{j}=so(N_{j})$ and
${\cal{P}}_{j}=(so(N_{j}))^{\perp}$ up to conjugacy. In a
decomposition {\bf{AII}} ${\cal{K}}_{j}=sp(N_{j}/2)$ and
${\cal{P}}_{j}=(sp(N_{j}/2))^{\perp}$ up to conjugacy.
\cite{Helgason}}. Denote by $\sigma_{j}$ $(S_{j})$ a generic element
of an orthogonal basis of $i{\cal K}_{j}$ ($i{\cal P}_{j}$). Let the
(total) Lie algebra $u(N_{1}N_{2}\cdot \cdot \cdot N_{n})$ be
decomposed as \be{OEdecom} iu(N_{1}N_{2}\cdot \cdot \cdot
N_{n})={\cal I}_{o} \oplus {\cal I}_{e}. \ee ${\cal I}_{o}$ (${\cal
I}_{e}$) is the vector space spanned by  matrices which are the
tensor products of an {\it odd} ({\it even}) number of elements of
the type $\sigma_{j}$. Then $u(N_{1}N_{2}\cdot \cdot \cdot
N_{n})=i{\cal I}_{o} \oplus i{\cal I}_{e}$ is a Cartan decomposition
i.e. \be{commIoIe} [i{\cal I}_o,i{\cal I}_o] \subseteq i{\cal I}_o,
\qquad [i{\cal I}_o,i{\cal I}_e] \subseteq i{\cal I}_e, \qquad
[i{\cal I}_e,i{\cal I}_e] \subseteq i{\cal I}_o. \ee \et The
decomposition (\ref{OEdecom}) is called a decomposition of the {\it
odd-even type}.

\vs

Associated to a Cartan decomposition (\ref{commIoIe}) is a Cartan
involution $\phi$ which is the identity on $i{\cal I}_{o}$ and
multiplication by $-1$ on $i{\cal I}_{e}$. A Cartan involution is
clearly a homomorphism. The structure of system (\ref{dinamica}) and
(\ref{dinamica1}) suggests that it is possible to choose this Cartan
involution as a homomorphism mapping the equations of two equivalent
models as in (\ref{condi11}). In fact, assume that there is   the
same number of subsystems (spin particles) in the two models and
that
 corresponding subsystems have the same dimension (namely the same
 spin). If we can display a decomposition
(\ref{63}) of the type {\bf{AI}} or {\bf{AII}} for every (spin)
 $su(N_j)$, such that $i\sigma_{x,y,z}\in {\cal K}_{j}$, then,
 for every value of the parameters,
 it holds that $B_{x,y,z}(') \in i{\cal I}_{o}$
and $A(') \in i{\cal I}_{e}$.  As
 shown in the following Theorems \ref{Teorema1}-\ref{Teorem3}
 decompositions of this type exist. We shall see in the following
 subsection that the Cartan involution associated to an odd-even
 type
 Cartan decomposition is the correct homomorphism to describe
 classes  of equivalent spin networks.  In fact,  not only models which
 are related by such a homomorphism are equivalent (according to
 Theorem \ref{THMGEN}) but the opposite is true as well. In other
 terms, two equivalent models are either exactly the same  or are related
 through such a homomorphism.

 \vs
The following three Theorems show the existence of a decomposition
of $su(N_j)$ of the type {\bf AI} or {\bf AII} where the subalgebra
${\cal K}_j$ contains the matrices $i\sigma_{x}$, $i\sigma_{y}$ and
$i\sigma_{z}$. Equivalently, they show the existence of a subalgebra
of
 $sp(\frac{N_{j}}{2})$ (type {\bf AII}) or $so(N_{j})$ (type {\bf AI})
 conjugate to the Lie algebra
 spanned by $i\sigma_{x}$, $i\sigma_{y}$ and $i\sigma_{z}$. The
 proofs are  presented in the following section.
 We shall see that the situation is different for integer and half
 integer spins.

 \bt{Teorema1} If the dimension $N_{j}$ of the system  is even (half
integer spin (Fermions)) there exists a subalgebra of
$sp(\frac{N_{j}}{2})$ conjugate to the Lie algebra spanned by
$i\sigma_x$, $i\sigma_y$ and $i\sigma_z$. \et \vs

\bt{Teorema2} If the dimension $N_{j}$ of the system  is odd (integer
spin (Bosons)) there exists a subalgebra of $so(N_{j})$ conjugate to
the Lie algebra spanned by $i\sigma_x$, $i\sigma_y$ and $i\sigma_z$.
\et

\vs

\bt{Teorem3} If the dimension $N_{j}$ of the system  is even (half
integer spin (Fermions)) there is no  subalgebra of $so(N_{j})$
conjugate to the Lie algebra spanned by $i\sigma_x$, $i\sigma_y$ and
$i\sigma_z$.
\et

\subsection{Necessary
and sufficient conditions for model equivalence}

In this subsection we will prove the equivalence result concerning
models of spin networks. This is given by  the following Theorem.


\bt{main}
 Let $\Sigma:= \Sigma(n,l_{j},J_{kl},\gamma_{k},\rho_{0})$
and $\Sigma':= \Sigma(n',l_{j}',J_{kl}',\gamma_{k}',\rho_{0}')$ be
two fixed models (see equations (\ref{dinamica}),
(\ref{dinamica1})). Assume that both models are controllable, that
 for model $\Sigma$ ($\Sigma'$), all the $\gamma_k$ ($\gamma_k'$)
 are non-zero and different {}from
each other, and that $\rho_0$ and $\rho_0'$ are not both scalar
matrices. Then
 $\Sigma$ is equivalent to $\Sigma'$ i.e.:
 \be{output-u}
 y_v(t):=Tr(S_{v}\rho(t))\equiv
 y'_v(t):=Tr(S'_{v}\rho'(t)), \ \text{ for } \ v\in \{x,\, y,\, z\},
 \ee
and for every control $u_x, u_y, u_z$, if and only if the following
condition holds:
\begin{quote}  {\bf{Cond. $(*)$}}: \ \
\begin{enumerate}

\item $n=n'$

Up to a permutation of the of the set $\{1,...,n\}$ (i.e. a
permutation of the indices for the particles)

\item $\gamma_k=\gamma_k'$,

and

\item
\be{nu12}
 l_k=l_k'.
 \ee

\item One of the following two conditions holds

\begin{enumerate}
\item
\be{uguali} A=A', \\ \text{and} \rho_0=\rho_0' \ee

\item Given the Cartan involution $\phi$ associated to the
decomposition of the odd-even type as {}from Theorem \ref{oeth}
\be{opposti} A'=\phi(A),
\\ \text{and}
i\rho_0'=\phi(i \rho_0)
\ee

\end{enumerate}
\end{enumerate}
\end{quote}

\et


The Theorem says that,  under appropriate controllability
assumptions, two equivalent models for spin networks are equivalent
if and only they have the same number of particles, corresponding
particles have the same spin, and their dynamical model and initial
state  are either the exactly the same or are related through the
Cartan involution associated to a decomposition of the odd-even
type. In  practical terms, given a general spin network, by driving
the network with an external electro-magnetic field and measuring
the total spin in the $x$, $y$ and $z$ direction, it is, in
principle, possible  to identify the number of particles, their
spin, the gyromagnetic ratios of every spin and the exchange
constants only up to a common sign factor, if the initial state is
not known. The proof that Condition ($*$) implies equivalence is an
application of the general property of Theorem \ref{THMGEN}. The
proof that equivalence implies Condition ($*$) is considerably
longer. However, several results can be obtained with proofs that
are formal modifications of the ones presented in \cite{LAA2005} for
the special case of spin $\frac{1}{2}$ particles. We shall  focus on
the new part of the proof needed to generalize to the case of
unknown spins.

\begin{center} {\em Condition ($*$) implies equivalence} \end{center}

It is clear that if
 (\ref{uguali}) holds, then the two models differ possibly only by a permutation
 of the indices of the particles. So they are equivalent.
 Assume now that Condition ($*$) holds with  (\ref{opposti})
 and assume for simplicity  (and without loss of generality)
 that the permutation of indices is the identity. Let $\phi$ be the Cartan
 involution associated to the decomposition of the odd-even type. We
 notice that
\be{nu112} \phi^{*}(iS_{v})=iS_{v}=iS'_{v}, \quad  v=x,y,z. \ee In
fact, given any $C\in u(N)$, we can  write $C=C_{o}+C_{e}$, with
$C_{o}\in i {\cal{I}}_{o}$ and $C_{e}\in i {\cal{I}}_{e}$, it holds:
\[
Tr(\phi^*(iS_v)C):=Tr((iS_v)\phi(C))=Tr((iS_v)(C_{o} -C_{e}))= Tr
((iS_v)C_{o})= Tr((iS_v)C),
\]
which,  since it has to hold for every $C$,  gives (\ref{nu112}).
Equations (\ref{opposti}) and (\ref{nu112})  imply that  equation
(\ref{condi11}) of Theorem \ref{THMGEN} holds. Since we also have
(\ref{statieq}), {}from (\ref{opposti}), we conclude that the two
models are equivalent  using Theorem \ref{THMGEN}.

\begin{center} {\em {Equivalence implies Condition (*)   }} \end{center}

The technique used in \cite{LAA2005} to prove this  result for
network of spin 1/2 particles extends   to  the general  case
treated here. However further analysis is required in this case,  in
particular to prove that equivalent spin networks have the same
values of the spins, while in \cite{LAA2005} it was assumed that the
networks were composed by all spin $\frac{1}{2}$'s. The main reason
why the proof in \cite{LAA2005} can be extended to this case is that
the basic commutation relations, which were the essential ingredient
of the proofs in \cite{LAA2005} still hold. More precisely,  the
matrices $\sigma_{x}$, $\sigma_{y}$ and $\sigma_{z}$ still satisfy,
for every value of the spin,  the commutation relations
(\ref{Commurel}).  This fact implies that it also holds: \be{p12}
\left[ I_{k_1v_{1},...,k_rv_{r}},
  I_{\bar{k}v_{\bar k}} \right]= \left\{ \begin{array}{cl}
                 0  & \text{if } \bar{k}\not\in
                 \{k_1,\ldots,k_r\}\\
                 0   & \text{if } \exists j \text{ with }\bar{k}=k_j
\text{ and }
                    v_{\bar{k}}=v_j \\
                 i I_{k_1v_{1},\ldots,k_j[v_jv_{\bar{k}}],\ldots,k_rv_{r}}
                & \text{if } \exists j \text{ with }\bar{k}=k_j \text
{ and }
                    v_{\bar{k}}\neq v_j
                     \end{array} \right. ,
\ee independently of the values of the spins.

\vs

Assume now that the two models $\Sigma$ and $\Sigma'$ are
equivalent. Then, using exactly the same arguments as in the proof
of Proposition 4.1 of \cite{LAA2005}, we obtain that the number of
the spin particle must be  the same, namely $n=n'$, and,  up to a
permutation of the indices, $\gamma_k=\gamma_k'$, $\forall k \in \{
1,...,n\}$, which is part 1. and 2. of Condition ($*$). Moreover as
in Proposition 4.1 of \cite{LAA2005}, we obtain \be{p2}
Tr(I_{kv}\rho(t))=Tr(I'_{kv} \rho'(t)), \\ \forall k \in \{ 1,...,n
\}, \\ \forall v \in \{ x,y,z \}. \ee Here $I'_{kv}$ is defined as
$I_{kv}$ but for $\Sigma'$ and, at this point, it may be different
{}from $I_{kv}$ since we have not shown yet that corresponding spins
must be equal. To prove this fact, we shall use Lemma \ref{plemma2}
below. The proof of this Lemma is a generalization of the proof of
Lemma 5.2 in \cite{LAA2005} where we use the general property
(\ref{nu4}) instead of the corresponding property for spin
$\frac{1}{2}$'s. We postpone this proof to Appendix A.

 \bl{plemma2} Assume that for all $t\geq 0$, all possible
trajectories $\rho(t)$ of $\Sigma$  and corresponding $\rho'(t)$ of
$\Sigma'$,  for fixed values $1\leq k_{1},\ldots,k_{r}\leq n$,
$v_{j}\in \{x,y,z\}$ and for given constants $\beta$ and $\beta'$,
we have: \be{p5} \beta Tr\left(I_{k_1v_1,...,k_r v_r}
\rho(t)\right)= \beta'
  Tr\left(I'_{k_1v_1,...,k_r v_r} \rho'(t)\right).
\ee
  Then
  \begin{enumerate}
      \item
  For any pair of indices $\bar k, \bar d\in \{1,\ldots,n\}$ with $\bar k
\in \{k_1,...,k_r\}$   and $\bar d \notin \{k_1,...,k_r\}$,
  \be{p6} \beta J_{\bar k \bar d}
Tr\left(I_{k_1v_1,...,k_r v_r, \bar d \bar v} \rho(t)\right)=
\beta'J_{\bar k \bar d}' Tr\left(I^{'}_{k_1v_1,...,k_r v_r,\bar d
\bar v } \rho'(t)\right), \ee
  for any value $\bar v \in \{x,y,z\}$.
\item For any pair of   indices $\bar k, \bar d$ both in
$\{k_1,...,k_r\}$, (for example $\bar k=k_1$, $\bar d=k_2$)  then
\be{p7} \beta (l_{\bar{d}}(l_{\bar{d}}+1))J_{\bar k \bar d}
Tr\left(I_{k_1v_1,k_3v_3,...,k_r v_r} \rho(t)\right)=
\beta'(l'_{\bar{d}}(l'_{\bar{d}}+1))J_{\bar k \bar d}'
Tr\left(I^{'}_{k_1v_1,k_3v_3,...,k_r v_r } \rho'(t)\right) . \ee
\end{enumerate}
  \el

In words, formula (\ref{p7})   means that  {}from (\ref{p5}), it is
possible to derive a new formula as follows. Select two indices in
the set $\{k_1,...,k_r\}$, $\bar k$ and $\bar d$. One of the two
indices (say $\bar d$) disappears {}from the subscript in the matrices
$I$ and corresponding $I'$. However a coefficient $l_{\bar
d}(l_{\bar d}+1)$ and $l^{'}_{\bar d}(l_{\bar d}+1)$ appears in the
left and right hand side, respectively, as well as a coefficient
$J_{\bar k \bar d}$ and $J^{'}_{\bar k \bar d}$.

We shall  now prove that, under the assumption of equivalence,  the
squares of the exchange constants $J_{dk}$ and $J'_{dk}$ must be
proportional, with a proportionality factor common to all pairs of
indices $d$ and $k$ and this will also be instrumental in the proof
of 3. of Condition $(*)$.

Fix any $1\leq k_1<k_2\leq n$, then, by applying statement $1.$ of
Lemma \ref{plemma2}, i.e. equation (\ref{p6}) with $\bar{k}=k_1$,
$\bar{d}=k_2$ to equation (\ref{p2}) with $k=k_1$, we have: \be{p17}
J_{k_1k_2} Tr\left( I_{k_1v_1,k_2v_2} \rho(t) \right) = J'_{k_1k_2}
Tr\left( I^{'}_{k_1v_1,k_2v_2} \rho'(t) \right), \ \  \ \forall\
v_1,v_2\in \{x,y,z\}. \ee Now, to the previous equality, we apply
statement $2.$ of Lemma \ref{plemma2}, i.e. equation (\ref{p7}) with
$\bar{k}=k_1$ and $\bar{d}=k_2$ to get:
\[
(l_{k_2}(l_{k_2}+1))J_{k_1k_2}^2 Tr\left( I_{k_1v_1} \rho(t) \right)
= (l'_{k_2}(l'_{k_2}+1))J^{'2}_{k_1k_2} Tr\left( I^{'}_{k_1v_1}
\rho'(t) \right),
\]
which, by equation (\ref{p2}), implies: \be{p18}
(l_{k_2}(l_{k_2}+1))J_{k_1k_2}^2=(l'_{k_2}(l'_{k_2}+1))J_{k_1k_2}^{'2}.
\ee Using the facts that the two indices $k_{1}$ and $k_{2}$ above
are arbitrary and that the graph  associated to the network  is
connected, by the controllability assumption, it is easy to see that
there exists  a positive constant $\alpha \in \RR$ such that, for
all $1 \leq  d < k \leq n$: \be{p18bis} J_{dk}^{2}= \alpha^{2}
{J'}_{dk}^2 \ \text{ and } \  l_{k}(l_{k}+1)=
\frac{1}{\alpha^{2}}l'_{k}(l'_{k}+1). \ee

\vs


Using (\ref{p18bis}), we can now prove $3.$ of Condition $(*)$. We
will do this using some lemmas and arguing by contradiction. First
notice that {}from (\ref{p18bis}), we have that if there exists
 a $\bar{k}\in \{1,\ldots,n\}$ such that $l_{\bar{k}}=l'_{\bar{k}}$,
then necessarily $\alpha^2=1$, thus $l_j=l'_j$ for all
$j=1,\ldots,n$, namely all the particles have the same spin. So if
we assume that (\ref{nu12}) does not hold, without loss of
generality, we can  assume $l_1>l'_1$. Using equation
(\ref{p18bis}), we get that $l_j > l'_j$ for all $j=1,\ldots,n$,
thus also $N_j>N'_j$. Let $R:=\frac{N}{N_1}=\prod_{j=2}^{n}N_{j}$ and
$R':=\frac{N'}{N'_1}=\prod_{j=2}^{n}N'_{j}$. \bl{111} For all $t \in
\RR$, and all the admissible trajectories $\rho$ and corresponding
trajectory $\rho'$, we have: \be{112}
\begin{array}{l} Tr\left((e^{i\sigma_z
t}\otimes {\bf 1}_{R\times R}) I_{1v} (e^{-i\sigma_z t}\otimes {\bf
1}_{R\times R}) \rho(s)
  \right) = \\
 Tr\left((e^{i\sigma_z t}\otimes {\bf 1}_{R'\times R'})I^{'}_{1v}
 (e^{-i\sigma_z t}\otimes
{\bf 1}_{R'\times R'} )\rho'(s)  \right),
\end{array}
 \ee
 for all $s\geq 0$.
 \el
 \bpr
First we notice that {}from  the Campbell-Baker-Hausdorff formula, we
have: \be{113} (e^{i\sigma_z t}\otimes {\bf 1}_{R\times R}) I_{1v}
(e^{-i\sigma_z t}\otimes {\bf 1}_{R\times R}) = \sum_{k=0}^{\infty}
\left( ad^{k}_{i\sigma_{z}\otimes {\bf 1}_{R\times R}} I_{1v}
\right) \frac{t^{k}}{k!}, \forall v \in \{x,y,z\}, \ee and an
analogous equation for $\Sigma'$. Moreover, by applying Lemma
\ref{lemma4} in Appendix A, with $W=I_{1v}$, $W'=I'_{1v}$, and
$k=1$, $v=z$, we have:
\[
Tr\left( ad_{i\sigma_{z}\otimes {\bf 1}_{R\times R}} I_{1v} \rho(s)
\right)= Tr\left( ad_{i\sigma_{z}\otimes {\bf 1}_{R'\times R'}}
I'_{1v} \rho'(s) \right).
\]
Now we can apply again Lemma \ref{lemma4} to the previous equality,
to get:
\[
Tr\left( ad^{2}_{i\sigma_{z}\otimes {\bf 1}_{R\times R}} I_{1v}
\rho(s) \right)= Tr\left( ad^{2}_{i\sigma_{z}\otimes {\bf
1}_{R'\times R'}} I_{1v}^{'} \rho'(s) \right).
\]
By applying repeatedly this procedure we obtain:
\[
Tr\left( ad^{k}_{i \sigma_{z}\otimes {\bf 1}_{R\times R}} I_{1v}
\rho(s) \right)= Tr\left( ad^{k}_{i\sigma_{z}\otimes {\bf
1}_{R'\times R'}} I_{1v}^{'} \rho'(s) \right),
\]
for all $k\geq 0$. Using this in  (\ref{113}), equation (\ref{112})
follows.
 \epr

 The proof of the following lemma is given in  Appendix A
  \bl{120}
The following formula holds:
 \be{121}
 (e^{i\sigma_z t}\otimes {\bf 1}_{R\times R})
I_{1x} (e^{-i\sigma_z t}\otimes {\bf 1}_{R\times R}) := P_{N_1}(t)
\otimes {\bf 1}_{R\times R} ,
 \ee
 where the matrix $P_{N_1}(\cdot)$ is periodic with period $2\pi$. Moreover
\be{122} P_{N_1}(\pi)=-P_{N_1}(0)=-\sigma_{x}. \ee
 \el

Using Lemmas \ref{111} and \ref{120}, we can now conclude the proof
that the spins are the same. Let $\bar{\rho}(s)\otimes {\bf
1}_{R\times R}$
 (resp. $\bar{\rho}'(s)\otimes {\bf 1}_{R'\times R'}$) the
orthogonal component of $\rho(s)$
 (resp. $\rho'(s)$) along $\sigma_x\otimes {\bf 1}_{R\times R}$ (resp.
 $\sigma_x\otimes {\bf 1}_{R'\times R'}$). Using equation
(\ref{121}), equality
  (\ref{112})  with $v=x$ can be written as :
  \be{131}
  Tr\left(P_{N_1}(t)\bar{\rho}(s)\right) R=
  Tr\left(P_{N'_1}(t)\bar{\rho}'(s)\right) R'.
  \ee
  Since we have assumed by contradiction $R>R'$, {}from (\ref{131})
  we have for every $t$:
\be{132}
  Tr\left(P_{N_1}(t)\bar{\rho}(s)\right)  <
  Tr\left(P_{N'_1}(t)\bar{\rho}'(s)\right).
  \ee
Now we will derive a contradiction by evaluating the previous
inequality at $t=0$ and $t=\pi$ and using (\ref{122}). In fact we
have:
\[
  Tr\left(P_{N_1}(0)\bar{\rho}(s)\right)  <
  Tr\left(P_{N'_1}(0)\bar{\rho}'(s)\right),
  \]
thus
\[
  Tr\left(P_{N_1}(\pi)\bar{\rho}(s)\right) =
   - Tr\left(P_{N_1}(0)\bar{\rho}(s)\right)  >
 - Tr\left(P_{N'_1}(0)\bar{\rho}'(s)\right)=
  Tr\left(P_{N'_1}(\pi)\bar{\rho}'(s)\right).
  \]
  The previous inequality contradicts equation (\ref{132}). Thus we
  conclude that $l_{1}=l'_{1}$, which implies that  equation
  (\ref{nu12}) holds.

\vs

 Since the two equivalent models $\Sigma$ and $\Sigma'$  have
  the same spin, the positive constant $\alpha$
  in (\ref{p18bis}) is equal to one. Therefore,
  for every pair $d,k \in \{ 1,...,n \}$,
   $J_{dk}$ and  $J_{dk}'$ only
   differ possibly by the a sign factor. Using the same argument as
   in the main Theorem of \cite{LAA2005} we can in fact conclude
   that there are only two possible case: The case where
   $J_{dk}=J^{'}_{dk}$ for every pair $d,k$ and the case where
   $J_{dk}=-J^{'}_{dk}$ for every pair $d,k$. If we are in the first case,
    then {}from the observability (which follows {}from controllability) of the model, we must have
$\rho_0=\rho_0'$, thus equation (\ref{uguali}) holds. This would be
the case  $a$ of part 4. of Condition $(*)$. On the other hand, if
$J_{kd}'=-J_{kd}$ for every pair $1\leq k<d \leq
  n$, we may conclude using Theorem \ref{THMGEN}. In fact we
  consider the  homomorphism $\phi$ given by the Cartan involution
  associated to the odd-even decomposition as in the previous part
  of the proof.  Conditions (\ref{condi11}) hold,
  thus, since the models are equivalent and observable, we get that:
  \[
  i\rho'_{0}=\phi (i\rho_{0}),
  \]
  thus equation (\ref{opposti}) holds.
  This concludes the proof of the Theorem.

\section{Proofs of Theorems \ref{Teorema1}-\ref{Teorem3}}
\label{Proofs}

 In the proofs of Theorems \ref{Teorema1} and \ref{Teorema2},
  we shall use the
following two types of elementary
 $k \times k$ matrices:

\be{CandT} C_k:=diag(-1,1,-1,...,(-1)^k), \qquad
T_k=adiag(1,1,1,...,1).   \ee The matrix $C_k$ is diagonal with
alternating elements while $T_k$ is antidiagonal with all ones on
the secondary diagonal and zeros everywhere else. Obvious properties
of these matrices are the following \be{Proper} C_k^2=T_k^2={\bf
1}_{k \times k}, \qquad T_k=T_k^T. \ee We are interested in the
action of these matrices by similarity transformation on diagonal
and tridiagonal $k \times k$ matrices. In particular, let us denote
by $D$ a generic, real,  diagonal,  $k \times k$ matrix and by $F$ a
generic, real, $k \times k$, tridiagonal matrix, which is also
symmetric and it has zero diagonal. If $M^a$ denotes the
antitransposed of $M$, namely the matrix obtained by reflecting
about the secondary diagonal, we can easily verify the following
properties.

\vs

\begin{enumerate}

\item

\be{Prop2} C_k DC_k=D, \qquad C_kFC_k=-F,  \ee

\item \be{Prop3}  T_kD T_k=D^a, \qquad T_k F T_k=F^a. \ee

\end{enumerate}

\vs

Now we are ready to prove Theorems \ref{Teorema1} and
\ref{Teorema2}.

\vs

{\bf Proof of Theorem \ref{Teorema1}} The matrices $i\sigma_z$ and
$i\sigma_x$ have (for every value of the spin) the following
structure \be{Strsz} i\sigma_z=i \pmatrix{D & 0 \cr 0 & -D^a},  \ee
\be{Strsx} i \sigma_x=i\pmatrix{F & P \cr P^T & F^a},  \ee where $F$
and $D$ have the structure above specified with $k:=\frac{N_j}{2}$
and $P$ is a $k \times k$ real matrix of all zeros except in the
$(k,1)$-th position. Now use \be{U} U:=\pmatrix{C_k & 0 \cr 0 &
T_k},  \ee which is orthogonal and therefore  unitary.

We calculate, using the first ones of (\ref{Prop2}) and
(\ref{Prop3})

\be{Trasz} i \tilde \sigma_z:= Ui\sigma_z U^*=i\pmatrix{D&0 \cr 0 &
-D}. \ee

Moreover, using the second ones of (\ref{Prop2}) and
(\ref{Prop3}), we have

\be{Trasx} i\tilde \sigma_x:=Ui \sigma_x U^*=i\pmatrix{-F & C_kPT_k
\cr T_k P C_k & F}.    \ee It is easily seen that $i \tilde
\sigma_z$ and $i \tilde \sigma_x$ are  symplectic, by  observing
that $C_kP T_k$ is a real symmetric matrix (only the $(k,k)$-th
element is different {}from zero). Therefore $sp(\frac{N_j}{2})$
contains a subalgebra conjugate to the one spanned by $i \sigma_x$
and $i\sigma_z$ and therefore $i\sigma_y$ and the Theorem is proved.
\qed

\vs

We now proceed to the proof of Theorem \ref{Teorema2}.

\vs

{\bf Proof of Theorem \ref{Teorema2}} In this case we set
$k:=\frac{N_j-1}{2}$. The matrix $i\sigma_z$ has the form \be{sz2}
i\sigma_z:= i \pmatrix{-D & 0 & 0 \cr 0 & 0 & 0 \cr 0 & 0 & D^a},
\ee with $D$ of dimension $k \times k$. Moreover $i \sigma_x$ has
the form \be{sx2} i \sigma_x:=i \pmatrix{F & v & 0 \cr v^T & 0 & w^t
\cr 0 & w &F^a}, \ee where $F$ is as above and $v$ ($w$) is a vector
of dimension $k$ with only the last (the first) component different
{}from zero, and the components different {}from zero are equal for $v$
and $w$. We use the unitary matrix \be{Unew} U:= \pmatrix{
\frac{i}{\sqrt{2}} C_k & 0 & (-1)^k \frac{i}{\sqrt{2}} T_k \cr 0 & 1
& 0 \cr \frac{1}{\sqrt{2}} T_k & 0 & \frac{1}{\sqrt{2}}C_k},  \ee
which is easily seen to be unitary by (\ref{CandT}) (\ref{Proper}).
We calculate. \be{sznew} i \tilde \sigma_z:= U i \sigma_z
U^*=i\pmatrix{ \frac{1}{2}(T_k D^a T_k- C_k D C_k) & 0 & \frac{i}{2}
( (-1)^k T_k D^a C_k -C_k D T_k ) \cr 0 & 0 & 0 \cr \frac{i}{2} (T_k
D C_k - (-1)^k C_k D^2 T_k) & 0 & \frac{1}{2} (C_k D^a C_k -T_k D
T_k) }. \ee Using the first ones of (\ref{Prop2}) and (\ref{Prop3}),
we find that  the diagonal blocks are zero. Moreover, the remaining
elements of the matrix are real so that $i \tilde \sigma_z$ is real.
Analogously, we calculate \be{sxnew} i\tilde \sigma_x:=U i \sigma_x
U^*=\ee $$ i\pmatrix{\frac{1}{2}(C_kF C_k + (-1)^{2k} T_k F^a T_k) &
 \frac{i}{\sqrt{2}} ( C_k v + (-1)^k T_k w) & \frac{i}{2}
 ( C_k F T_k + (-1)^k T_k F^a C_k) \cr
 * & 0 & \frac{1}{\sqrt{2}} ( v^T T_k+ w^T C_k) \cr
 * & * & \frac{1}{2}( T_k F T_k+ C_k F^aC_k)},
 $$
where we have denoted by a star $*$ the components that can be
obtained {}from the requirement that the matrix is skew Hermitian. Now
the $(1,1)$ and $(3,3)$ blocks are zero {}from the second ones of
properties (\ref{Prop2}) and (\ref{Prop3}), while the (2,3) block is
zero because of the structure  of the vectors $v$ and $w$. All the
other blocks are purely real matrices so that $i \tilde \sigma_x$ is
also in $so(n)$ and this completes the proof. \qed

\vs

We now give the proof of the negative result in Theorem
\ref{Teorem3}.

{\bf Proof of Theorem \ref{Teorem3}}  Assume that there exists a
matrix $X \in SU(N_j)$ such that \be{T31} X i\sigma_x X^* :=\tilde
R_x, \ee \be{T32} X i \sigma_y X^* :=\tilde R_y, \ee \be{T33} X
 i \sigma_z X^* :=\tilde R_z, \ee with $\tilde R_x,$ $\tilde R_y$ and
$\tilde R_z$ in $so(N_j)$. Then we can use the {\bf AI} Cartan
decomposition of $SU(N_j)$ \cite{Helgason} to write $X$ as \be{A9}
X=K_1 A K_2, \ee with $K_1$ and $K_2$ in $SO(N_j)$ and $A$ diagonal
i.e. \be{FormA} A:=diag(e^{i \phi_1},...,e^{i\phi_{\frac{N_j}{2}
}}). \ee Therefore we can write \be{A10} K_1 A K_2 i \sigma_{x,y,z}
K_2^T \bar A K_1^T= \tilde R_{x,y,z}, \ee or, defining
$R_{x,y,z}:=K_1^T \tilde R_{x,y,z}K_1$ which is also
 real skew-symmetric, we can write
\be{For1}  K_2 i \sigma_{x,y,z} K_2^T=\bar A R_{x,y,z} A.   \ee

\vs

The real matrices $R_{x,y,z}$ must satisfy the same basic
commutation relations (\ref{Commurel}) of $i\sigma_x$, $i \sigma_y$
and $i \sigma_z$ and have the same eigenvalues of $i\sigma_x$,
$i\sigma_y$ and $i\sigma_z$, namely for a (half integer) spin $j$,
$\pm ji$, $\pm (j+1)i$,...,$\pm \frac{1}{2}i$. We now study the
structure of $R_{x,y,z}$ in (\ref{For1}) and get a contradiction
with these facts.

\vs

First notice that, since $A$ is diagonal as in (\ref{FormA}), the
action of $A$ on the right hand side of (\ref{For1}) namely, $R
\rightarrow \bar A R A$ changes the (real) element $r_{jk}$ of $R$
into $r_{jk}e^{-i(\phi_j- \phi_k)}$. Since the entries on the left
hand side of (\ref{For1}) are either all purely imaginary or all
purely real, if $\phi_j-\phi_k$ is not a multiple of $\frac{\pi}{2}$
then we must have $r_{jk}=0$. Consider the indices $1,...,N_j$ and
let ${\cal O}$ be the set of indices $k$ such that $\phi_1-\phi_k$
is and odd multiple of $\frac{\pi}{2}$ and ${\cal E}$ the set of
indices $k$ such that $\phi_1-\phi_k$ is an even multiple of
$\frac{\pi}{2}$, and $\cal N$ the set of indices $k$ such that
$\phi_1- \phi_k$ is not an integer  multiple of $\frac{\pi}{2}$


 {}From
(\ref{For1}) it follows that since $i \sigma_y$ is real the terms
$r_{jk}$ of $R_y$ where $j$ and $k$ belong to different sets must be
zero because in that case $e^{i (\phi_j - \phi_k)}$ in (\ref{FormA})
has a nonzero imaginary part  in this case. Therefore only the
elements $r_{jk}$ where $j$ and $k$ belongs both to ${\cal O}$, or
both to $\cal E$, or both to $\cal N$, are possibly different {}from
zero. Therefore after possibly a reordering of rows and columns,
which corresponds to a similarity transformation by a permutation
matrix, $R_y$ must be of block diagonal form and without loss of
generality and for simplicity we shall assume only two blocks
(rather than three). Therefore we write   \be{RY}
R_y:=\pmatrix{Y_{11} & 0 \cr 0 &Y_{22}},
 \ee
where $Y_{11}$ has dimensions $n _o \times n_o$ with $n_o$ the
cardinality of $\cal O$ and $Y_{22}$ has dimension $(n-n_o) \times
(n-n_o)$. Both $Y_{11}$ and $Y_{22}$ are skew-symmetric matrices. An
analogous argument shows that, after possibly the same re-ordering
of column and row indices, $R_z$ can  be written as \be{RZ}
R_z:\pmatrix{0 & Z_{12} \cr - Z_{12}^T & 0},  \ee where $Z_{12}$ is
a general matrix of dimensions $n_o \times (n -n_o)$.


Now consider the possible values for $n_o$. $n_o$ odd is to be
excluded because this would cause $det Y_{11} =0$ in (\ref{RY}) and
this contradicts the fact that $R_y$ has no zero eigenvalues.
Moreover $n_o \not= (n-n_o)$ (i.e. $n_o \not= \frac{N_j}{2}$) would
cause $R_z$ to have determinant equal to zero. This can  be easily
verified by calculating \be{detrz} det(R_z^2)=(det
R_z)^2=det\pmatrix{ -Z_{12} Z_{12}^T & 0 \cr 0 & -Z_{12}^T Z_{12}},
\ee since, in this case,  at least one of the matrices on the
diagonal blocks does not have full rank. These considerations
already exclude the cases where $\frac{N_j}{2}$ is an odd number as
for spins $\frac{1}{2}$, $\frac{5}{2}$, $\frac{9}{2}$ etc. and we
can assume $R_y$ and $R_z$ of the form (\ref{RY}) and (\ref{RZ})
with $n_o=\frac{N_j}{2}$. To obtain a contradiction in this case too
we first notice that, since $Y_{11}$ and $Y_{22}$ have even
dimension and are skew-symmetric, we can apply a similarity
transformation $T:=\pmatrix{T_1 & 0 \cr 0 & T_2}$, with $T_1$ and
$T_2$ orthogonal so that $TR_yT^T$ is block diagonal \be{TRY}
TR_yT^T=(D_1,D_2,...,D_{\frac{N_j}{2}}), \ee where the $2 \times 2$
block $D_k$ has the form \be{DKKKK} D_k:=\pmatrix{0 &
 l_k \cr -l_k &0}, \ee
where each $l_k$ corresponds to a couple of complex conjugate
eigenvalues of $R_y$ so that $l_k=\frac{p}{2}$ with $p$ odd
corresponds to the pair $\pm \frac{p}{2}i$. Moreover we choose $T$
so that the first $\frac{N_j}{4}$ blocks are ordered according to
the increasing value of $l_k$ and the same thing for the last
$\frac{N_j}{4}$ blocks. We shall therefore assume this structure of
$R_y$ in the remainder of the proof. We notice also that the
transformation $TR_zT^T$ does not change the structure of $R_z$ as
$Z_{12}$ in (\ref{RZ}) was chosen to be a general $\frac{N_j}{2}
\times \frac{N_j}{2}$ real matrix. Express  $Z_{12}$ in terms of $2
\times 2$ blocks $L_{fk}$, $f,k=1,...,\frac{N_j}{4}$,
$k=\frac{N_j}{4} +1,..., \frac{N_j}{2}$, which is possible since
$\frac{N_j}{2}$ is an even number. Now, we impose the fact that
$R_y$ and $R_z$ have to satisfy the same commutation relations as
$i\sigma_y$ and $i\sigma_z$. In particular, we must have
\be{commmmm} [[R_y,R_z], R_y]=R_z.  \ee This equation gives for the
$L_{fk}$ block the following one \be{blp} 2D_f L_{fk}
D_k-L_{fk}D_k^2-D_f^2 L_{fk}= L_{fk}.  \ee If we write the generic
$L_{fk}$ as \be{forljk} L_{fk}:=\pmatrix{a_1 & a_2 \cr a_3 & a_4},
\ee and recall the structure of $D_f$ and $D_k$, \be{DjDk}
D_f:=\pmatrix{0 & l_f \cr -l_f & 0}, \qquad D_k:=\pmatrix{0 & l_k
\cr -l_k &0},  \ee we obtain the following equations for $a_2$ and
$a_3$ (and analogous equations for $a_1$ and $a_4$) \be{ea2a31} 2l_k
l_f a_3=(1-l_f^2- l_k^2) a_2, \ee \be{a2a32} 2l_k l_f a_2=(1-l_f^2
-l_k^2)a_3.  \ee Combining these, we obtain \be{comb} 4l_k^2l_f^2
a_3=(1-l_k^2-l_f^2)^2 a_3, \ee which shows, taking the square root
of both sides, that the only possibilities to have $a_3$ and
therefore $a_2$ different {}from zero are the cases $l_f +l_k=\pm 1$.
In these cases we can easily see that \be{jppo} a_3=-a_2. \ee
Similarly, one finds that we have $a_4$ and $a_1$ in (\ref{forljk})
different {}from zero if and only if $l_f+l_k=\pm 1$ and, in these
cases we have \be{jppo2} a_1=a_4.  \ee In conclusion, all the blocks
$L_{fk}$ are zero except the ones corresponding to indices $f$ and
$k$ with neighboring values of $l_f$ and $l_k$ which have the
structure \be{stry} L_{fk}:=\pmatrix{x & y \cr -y & x}. \ee
Therefore $R_z$ has the form in (\ref{RZ}) where the $f-$th block
row of $Z_{12}$ has at most two blocks different {}from zero and with
the structure in (\ref{stry}). We denote these blocks by $P_f$ and
$S_f$, where $P$ ($S$) stands for 'predecessor' ('successor') and
correspond to the index $k$ such that $l_k=l_f-1$ and $l_k=l_f+1$,
respectively. Now, we argue that a matrix $R_z$ with this structure
must necessarily have all the (purely imaginary) eigenvalues with
multiplicity at least  two and this gives the desired contradiction
because $R_z$ should have the same spectrum of $i\sigma_{x,y,z}$
which consists of all simple eigenvalues. In order to see this fact,
re-consider the block structure of $R_y$ in (\ref{TRY}). If the
blocks corresponding to eigenvalues $\pm \frac{1}{2}i$ and $\pm
\frac{3}{2}i$ belong to the same half, then the corresponding matrix
$R_z$ will have a two dimensional block row (or column) equal to
zero,  and therefore $0$ will be an eigenvalue with multiplicity at
least $2$. Therefore we can assume that these two blocks belong to
two different halves and by the ordering we have imposed they must
be the first ones of each half. Assume that the block corresponding
to $\pm \frac{1}{2}i$ is in the first half. If this is not the case,
consider the transposed of $R_z$ and repeat the arguments that
follow. It is possible to choose a block diagonal similarity
transformation \be{simil}
U:=diag(G_1,G_2,...,G_{\frac{N_j}{4}},F_{1},F_2,...,F_{\frac{N_j}{4}}),
\ee with all the $G_f$'s and $F_f$'s $2 \times 2$ orthogonal
matrices so that $UR_zU^T$ has the same structure as before but all
the matrices $P_j$ and $S_j$ are {\it scalar} matrices. We construct
the matrix $U$ proceeding by block rows. The first block row
contains only $S_1$ as $\frac{1}{2}$ has no predecessors. All the
zero blocks remain zero and $S_1$ is transformed as \be{S1transf}
G_1S_1F_1^T. \ee We choose $F_1={\bf 1}_{2 \times 2}$ and $G_1$,
which has the general form \be{G1} G_1:=\pmatrix{cos(\theta) &
sin(\theta) \cr -sin(\theta) & cos(\theta)},  \ee so that
$sin(\theta)x +cos(\theta)y=0$ if $S_1:=\pmatrix{x & y \cr -y & x}$.
This will give a scalar matrix. At the generic $f-$th block row,  we
have, at the most two nonzero blocks $P_{fk}$ and $S_{fb}$ where we
now use an extra index $k$ and $b$ to indicate the block column to
which they belong. They transform as \be{tran1} P_{fk} \rightarrow
G_f P_{fk} F_k^T,  \ee and \be{tran2} S_{fb} \rightarrow G_f
S_{fb}F_b^T, \ee respectively, while all the other blocks remain
zero. If $F_k$ has not been chosen before we set $F_k={\bf 1}_{2
\times 2}$. In any case, we choose $G_f$ as before to make $G_f
P_{fk} F_k^T$ a scalar matrix. We then choose $F_b$ to make $G_f
S_{fb}F_b^T$ a scalar matrix. $G_f$ and $F_b$ had not been chosen at
previous steps. This is obvious for $G_f$ and follows by an
induction argument for $F_b$ since all the $F$ matrices chosen
before the $f-$th step correspond to predecessors and successors
with (column) indices strictly less then $b$ (recall that  in the
two halves of the matrix $R_y$ the blocks are arranged in increasing
order of (absolute value of) eigenvalue). In conclusion, modulo  the
similarity transformation defined by $U$ in (\ref{simil}), we can
assume that $R_z$ has the form \be{conclu} R_z=K \otimes I_{2 \times
2},  \ee where $K$ is a skew-symmetric $\frac{N_j}{2} \times
\frac{N_j}{2}$ matrix. By known results on the eigenvalues of the
Kronecker products of two matrices, it follows that the eigenvalues
of $R_z$ are the same as those of $K$ each with multiplicity at
least two. This gives the desired contradiction and concludes the
proof of the Theorem. \qed

\section{Conclusions}
\label{CONCLU}

This paper has presented a collection of mathematical results
concerning the input-output equivalence of quantum systems. Models
that are equivalent cannot be distinguished by an external observer
and therefore the determination of parameters in a quantum
Hamiltonian can only be obtained up to equivalent models. Motivated
by recent results on the isospectrality of quantum Hamiltonians
\cite{Luban} in molecular magnets, we have completely characterized
the classes  of spin networks which are equivalent. In several
cases,  the characterization of equivalent models can be obtained
through a Lie algebra homomorphism which is suggested by a Cartan
structure of the underlying dynamics.

We believe many of the results and the concepts presented in this
paper for quantum systems  could be generalized to classes of
systems
 relevant in other applications with both dynamics and output
 linear in the state. This will be the subject of further research.


\vs

\vs

{\bf Acknowledgment} D. D'Alessandro research was supported by NSF
under Career grant ECS-0237925.

\section*{Appendix A: Additional results and proofs}

The proof of the following Lemma can be obtained with a formal
modification of the proof of Lemma 4.4 in \cite{LAA2005} and it is
therefore omitted.

\bl{lemma4} Let  $\Sigma$  and $\Sigma'$ be two equivalent models
models.  If $W$ and $W'$ are two given Hermitian matrices such that
\be{C10} Tr(W\rho(t))=Tr(W'\rho'(t)), \ee for every pair of
corresponding trajectories $\rho(t)$ and $\rho'(t)$, then it also
holds \be{C11}
Tr([W,I_{kv}]\rho(t))=Tr([W',I^{'}_{{(k)}v}]\rho'(t)), \ \ \forall \
k\in\{1,\ldots,n\}, \forall v \in \{x,y,z\},\ee up to a permutation
of the indices \footnote{This permutation is the same and fixed for
all the results where it is mentioned}.\el

\subsection*{Proof of Lemma \ref{plemma2}}

We first state a Lemma whose proof can be obtained {}from the proof of
Lemma 5.1 in \cite{LAA2005} and then proceed to the proof of Lemma
\ref{plemma2}

\bl{plemma1} Assume that for all $t\geq 0$, all the possible
trajectories $\rho(t)$ of  $\Sigma$
 and corresponding $\rho'(t)$ of  $\Sigma'$, for
fixed values $1\leq k_{1},\ldots, k_{r}\leq n$, and fixed  $v_{j}\in
\{x,y,z\}$, we have: \be{p3} Tr\left(I_{k_1v_1,...,k_r v_r}
\rho(t)\right)=
  Tr\left(I^{'}_{k_1v_1,...,k_r v_r} \rho'(t)\right).
\ee
  Then:
  \begin{enumerate}
      \item
  equation (\ref{p3}) holds for any possible choice of the values of
  $v_{j}\in \{x,y,z\}$;
  \item
  \be{p4}
  \begin{array}{l}
  Tr\left(\left[ \left[iI_{\bar d v_{\bar{d}}},[iI_{\bar k
  v_{\bar{k}}},A]\right], I_{k_1v_{1},...,k_rv_{r}}\right]
  \rho(t)\right)= \\
Tr\left(\left[ \left[iI^{'}_{\bar d v_{\bar{d}}},[iI^{'}_{\bar k
  v_{\bar{k}}},A']\right], I^{'}_{k_1v_{1},...,k_rv_{r}}\right]
  \rho'(t)\right),
  \end{array}
\ee
  for all the indices $1\leq \bar{d}\neq \bar{k}\leq n$ and every
 $\{v_{\bar{d}}\neq v_{\bar{k}}\}\in \{x,y,z\}$.
  \end{enumerate}
\el

\vs

We now proceed to the proof of Lemma \ref{plemma2}.
First notice that  from Lemma \ref{plemma1}, it is enough to prove
(\ref{p6}) and (\ref{p7}) for a particular choice of $\{v_j\}$ and
$\bar{v}$. Moreover, we have, for $\bar{d}>\bar{k}$, \be{p15} \left[
iI_{\bar{d}z},\left[iI_{\bar{k}x}, A\right]\right] =
-J_{\bar{k}\bar{d}}i I_{\bar{k}z,\bar{d}x}. \ee

1. By applying Lemma \ref{plemma1} (equation (\ref{p4})) to
(\ref{p5}) and using (\ref{p15}) we get: \be{p16} \beta
  Tr\left(\left[-J_{\bar{k}\bar{d}}i I_{\bar{k}z,\bar{d}x},
I_{k_1v_1,...,k_r
  v_r}\right]
\rho(t)\right)=
  \beta^{'}
  Tr\left(\left[-J'_{\bar{k}\bar{d}}i I^{'}_{\bar{k}z,\bar{d}x},
I^{'}_{k_1v_1,...,k_r
  v_r}\right]
\rho'(t)\right). \ee We may assume, without loss of generality, that
$\bar{k}=k_j$ and $v_j=x$. In this case we have:
\[
-J_{\bar{k}\bar{d}} \left[ I_{\bar{k}z,\bar{d}x}, I_{k_1v_1,...,k_r
  v_r}\right]=J_{\bar{k}\bar{d}} i
  I_{k_1v_1,\ldots,k_jy,\ldots,k_rv_r,\bar{d}x}.
  \]
  Combining the previous equality with (\ref{p16}), equation
  (\ref{p6}) follows easily.

  2. Using the same procedure, we obtain again equation
  (\ref{p16}), but now both indices $\bar{k}$ and $\bar{d}$ are in
  $\{k_1,\ldots,k_r\}$. Assume, for example that $k_1=\bar{k}$ and
  $k_2=\bar{d}$, and take $v_{k_1}=v_{k_2}=x$.

 Now we have:
  \be{nu}
\left[ I_{k_1z,k_2x}, I_{k_1x,k_2x,\ldots,k_r
  v_r}\right]=
  I_{k_1y,k_2x^2,k_3v_3\ldots, k_rv_r},
  \ee
where, with this notation, we mean that in the $k_2^{th}$ position we
have the matrix ${\sigma}{_x^2}$. Thus, combining equations
(\ref{p16}) and (\ref{nu}), we get: \be{nu1}
 \beta J_{k_1k_2}
  Tr\left(I_{k_1y,k_2x^2,k_3v_3\ldots, k_rv_r}
\rho(t)\right)=
  \beta' J'_{k_1k_2}
  Tr\left(I^{'}_{k_1y,k_2x^2,k_3v_3\ldots, k_rv_r}
\rho'(t)\right). \ee

Using the same procedure, we  conclude: \be{nu2}
  \beta J_{k_1k_2}
  Tr\left(I_{k_1y,k_2y^2,k_3v_3\ldots, k_rv_r}
\rho(t)\right)=
  \beta' J'_{k_1k_2}
  Tr\left(I^{'}_{k_1y,k_2y^2,k_3v_3\ldots, k_rv_r}
\rho'(t)\right), \ee and  \be{nu3}
  \beta J_{k_1k_2}
  Tr\left(I_{k_1y,k_2z^2,k_3v_3\ldots, k_rv_r}
\rho(t)\right)=
  \beta' J'_{k_1k_2}
  Tr\left(I^{'}_{k_1y,k_2z^2,k_3v_3\ldots, k_rv_r}
\rho'(t)\right).  \ee

 Adding together  equations (\ref{nu1}), (\ref{nu2}), and (\ref{nu3}) and
using  (\ref{nu4}), we get:
\[
\beta (l_{k_2}(l_{k_2}+1))J_{k_1k_2}
  Tr\left(I_{k_1y,k_3v_3\ldots, k_rv_r}
\rho(t)\right)=
  \beta' (l_{k_2}'(l_{k_2}'+1))J'_{k_1k_2}
  Tr\left(I^{'}_{k_1y,k_3v_3\ldots, k_rv_r}
\rho'(t)\right),
\]
as desired.

\subsection*{Proof of Lemma \ref{120}}

\bpr

 We recall the formulas (\ref{121}) and (\ref{122}) to
 be proved, i.e.:
 \be{aa121}
 (e^{i\sigma_z t}\otimes {\bf 1}_{R\times R})
I_{1x} (e^{-i\sigma_z t}\otimes {\bf 1}_{R\times R}) := P_{N_1}(t)
\otimes {\bf 1}_{R \times R},
 \ee
 where the matrix $P(\cdot)$ is periodic with period $2\pi$, and
\be{aa122} P_{N_1}(\pi)=-P_{N_1}(0)=\sigma_{x}. \ee The proof can be
done directly by computing the matrix above. This is simplified by
the fact that the matrix $\sigma_{z}$ is  always a diagonal matrix.
We will give an outline of the argument when $l_{1}$ is half integer
spin. The idea is to use the representations for the matrices
$\sigma_{z}$ and $\sigma_{x}$ given by equations (\ref{Strsz}) and
(\ref{Strsx}). The case of integer spin can be derived similarly
starting with the representations given by equations (\ref{sz2}) and
(\ref{sx2}).

Using equations (\ref{Strsz}) and (\ref{Strsx}), we obtain
\be{aaexp}
 e^{i\sigma_z t}i\sigma_{x}
e^{-i\sigma_z t} = \pmatrix{e^{iDt}Fe^{-iDt} & e^{iDt}Pe^{iD^{a}t}
\cr e^{iD^{a}t}P^Te^{-iDt} & e^{-iD^{a}t}F^ae^{iD^{a}t}}. \ee The
properties of the matrices $D$, $P$ and $F$ are  described
 in Section \ref{Proofs}. Moreover $D=diag(j,j-1,...,\frac{1}{2})$
 for a half integer spin $j$. By using these properties,
 it follows that all the time
 depending terms in equation (\ref{aaexp}) are of the form
  $e^{it}$.  Thus matrix (\ref{aaexp}) is periodic of period $2\pi$.
  The fact that the dependence is of the type $e^{it}$,
  in turn, implies that equations (\ref{aa121}) and (\ref{aa122})
 hold.

\epr

\end{document}